# Novel single phase vanadium dioxide nanostructured films for methane sensing near room temperature

A. K. Prasad[1*], S. Amirthapandian[1], S. Dhara[1], S. Dash[1], N. Murali[2], A.K. Tyagi[1]

[1] Materials Science Group, Indira Gandhi Centre for Atomic Research, Kalpakkam, TN 603102, India

* Email: akp@igcar.gov.in

[3] Electronics and Instrumentation Group, Indira Gandhi Centre for Atomic Research, Kalpakkam, TN 603102, India

**Abstract:** Methane ($CH_4$) gas sensing properties of novel vanadium dioxide ($VO_2$) nanostructured films is reported for the first time. The single phase nanostructures are synthesized by pulsed dc-magnetron sputtering of V target followed by oxidation in $O_2$ atmosphere at 550°C. The partial pressure of $O_2$ is controlled to obtain stoichiometric $VO_2$ with the samples showing rutile monoclinic crystalline symmetry and regions of rod shaped nano-architectures. These nanostructured films exhibit a reversible semiconductor to metal transition in the temperature range of 60-70°C. Gas sensing experiments are carried out in the temperature span from 25°C to 200°C in presence of $CH_4$. These experiments reveal that the films respond very well at temperatures as low as 50°C, in the semiconducting state.

*Corresponding Author, Email: akp@igcar.gov.in

Fax: +91 – 044- 27480081, Ph: +91-044-27480500 Ext 21512

## 1. Introduction

Detection of methane ($CH_4$) using semiconducting metal oxides below 200°C is known to be difficult due to its thermodynamic stability. Efforts to reduce the operating sensing temperature have focused on addition of nanoclusters of noble metals or transition elements to metal oxides. Most of the $SnO_2$ based sensors detect methane at higher temperatures (>300°C) [1]. The lowest operating temperature for detecting $CH_4$ using metal loaded $SnO_2$ is 220°C [1, 2] and for loaded ZnO is 75°C [3]. But the detection limit is in percentage or thousands of ppm at the least and sensor response value, defined as $(R_{air}-R_g)/R_{air}$, was in the range of 5-15%. Apart from metal oxides, Pd/MWCNT nanocomposites have been used recently to detect $CH_4$ at room temperature (RT) [4, 5]. The highest value of S obtained for 3% concentration of $CH_4$ was around 3%.

In this regard, the potential of nanostructured vanadium oxide materials in gas sensing application is explored. Though there are numerous reports on $V_2O_5$ as a gas sensing material, till date, there has been very little interest shown towards $VO_2$ for gas sensing. Rella et al. [6] observed the response of the mixture of $V_2O_5$ and $VO_2$ towards $NO_2$ and ethanol. Recently, there have been reports that deal with $H_2$ detection of undoped and Pd functionalized $VO_2$ [7, 8]. The rationale behind using $VO_2$ is that it is a semiconducting oxide at RT with relatively lower resistance (in few 10s of kΩ) when compared to other metal oxides. Moreover, $VO_2$ has recently attracted widespread attention due to its semiconductor to metal transition (SMT) around 68°C [9]. Such an SMT is promising in terms of applications in optical switching devices, thermal relays, energy management devices, infrared sensors and optical recording devices [10, 11].

Preparation of stoichiometric $VO_2$ requires precise control of deposition parameters due to the possibility of formation of other thermodynamically stable oxides like $V_2O_3$ and $V_2O_5$. During the past decade, $VO_2$ has been synthesized by various techniques such as sol-gel [12], sputtering [12-14], pulsed laser deposition [9, 15], chemical vapor deposition (CVD) and hydrothermal synthesis [16].

We have adopted a two-step synthesis approach using pulsed dc magnetron sputtering and post-deposition oxidation to obtain nanostructured $VO_2$ films with desired composition and morphology. This technique yields films of higher film density with better adhesion due to higher ion energy when compared to hydrothermal and other vapor phase deposition techniques. Using $VO_2$ thin films, $CH_4$ detection close at around 50°C has been shown. This is the first report on $VO_2$ gas sensing response towards $CH_4$, to the best of our knowledge.

## 2. Experimental

### 2.1. Synthesis of Films

Nanostructured films of metallic vanadium were deposited on sensor substrates (Au interdigitated electrodes pre-patterned on alumina) using pulsed dc magnetron sputtering technique. The pulse power was 160 W at 100 kHz and base pressure was $3 \times 10^{-6}$ Torr using $Ar^+$ plasma. The substrate temperature was maintained at 350°C during the deposition time of 40 min. These samples were then oxidized in an annealing chamber at 550°C in $O_2$ environment where the flow was maintained at 20 sccm at a partial pressure of 0.2 Torr. The oxidation was performed for 3 hours to achieve an oxide layer thickness of 450 nm.

## 2.2. Characterization

The morphology of these samples were studied by field emission scanning electron microscope (FESEM; Zeiss Supra 55) operated in secondary electron imaging mode. The crystal structure and orientation of the nanostructured films were determined by grazing incidence X-ray diffractometer (GIXRD; Bruker D8 Discover) equipped with monochromatic Cu-K$\alpha$1 X-ray source($\lambda$ = 1.5406 Å). The spectrum was obtained at diffraction angle (2$\theta$) ranging from 15° to 60° at a grazing angle of 1.5°. The crystallite orientation, grain sizes and structure were further confirmed using high resolution transmission electron microscope (HRTEM; Zeiss Libra 200 FE) operated at 200 kV. Temperature dependent resistance measurements were obtained from Linkam hot stage with precise temperature control ($\pm$0.1°C) coupled with Agilent B2902A Precision Source/Measurement Unit.

## 2.3. Gas sensing

The gas sensing experiments were carried out in a custom built exposure facility system [17]. A brief description is given here. It consists of a double walled stainless steel chamber containing a PID controlled hot stage upon which the sample is mounted. The gas flows are set from mass flow controllers and the sensor response upon exposure to gases is manifested as resistance changes, which is recorded in a standard two probe method using a multimeter. Commercially procured gas mixtures of ultra high pure grade $CH_4$, $NH_3$, $H_2$ and $NO_2$ diluted in nitrogen carrier gas were used along with synthetic air background. These gases were further passed through moisture traps to remove any residual humidity. The concentrations were varied in the range from 50 - 500 ppm at operating temperatures of 25, 50, 100, 150 and 200°C. The sensor response (S) expressed in % is defined as ratio of change in resistance upon exposure to gas to the initial resistance ($\Delta R/R_0$) ×100 where $\Delta R = (R_0 – R_{gas})$, $R_{gas}$ is the resistance upon exposure to gas and $R_0$ is the initial resistance.

## 3. Results and Discussions

The morphology of the films determined using FESEM are shown in Figs. 1a&1b. The films are dense with several regions consisting of nanorods of 100-150 nm in thickness and several micrometers in length appearing in bunch (Fig. 1b). The layered structure present in the nanorod bundles is evident from Fig. 1a. The bundled growth may be due to the presence of strong Van der Waals force of interaction operating among nearby nanorods, in the layered structure, holding them together in a similar manner as explained earlier by Sun et al. [18].

The GIXRD study shows peaks corresponding to ($\bar{2}01$), (201), (021), (002), ($\bar{2}02$) and (222) planes (Fig. 1c) which confirm the presence of single phase polycrystalline $VO_2$. The peaks match well with the monoclinic rutile [JCPDS file 33-1441] phase of $VO_2$ having space group C2/m. The substrate peaks are marked by '*'. It is observed that the peaks of {00$l$} family correspond to the layered structure as is evident from the presence of peak corresponding to (002) [19]. The occurrence of (222) family of planes can be attributed to the oriented nature of growth. The {$h0l$} family of planes indicates the presence of two dimensional nanostructures in the $VO_2$ films [20].

These nanorods are further observed by HRTEM. The nanorods are 100-150 nm in thickness (Fig. 2a) confirming the findings from FESEM study (Fig. 1a). The appearance of bundles is also evident from the brightness contrast of the nanorods in the image. The selected area electron diffraction (SAED) pattern of the tip of the bundle (Fig. 2b) shows the presence of (002) and (201) reflections corroborating the observation already made from GIXRD studies. The zone axes are found lying along [010]. Fig. 2c shows the HRTEM image of the tip of the bundle which depicts a grain having (201) orientation with a *d* spacing of 3.17 Å.

$VO_2$ is distinct in its characteristic semiconductor to metal transition as is observed in Fig. 3. It shows transition around 68°C during heating and 58°C during cooling. The resistance is reversible as evident from the graph with hysteresis loop width of around 10°C. It was observed that the resistance changed more than two orders of magnitude over the transition region. A sharp change across the transition temperature shows presence of stoichiometric $VO_2$.

The sensing behavior at three different temperatures viz. 50, 100 and 150°C for 50-500 ppm $CH_4$ are plotted (Fig. 4). There is hardly any response to $CH_4$ at 25°C and above 150°C. It can be observed from the resistance transients (Fig. 4a) that upon exposure to $CH_4$, the resistance of the $VO_2$ sensor decreases. As observed from the slopes of the transient plots, it can be seen that the response and recovery times are very short for sensing performed at 50°C in its semiconducting state. The response and recovery times increase as the temperature is raised. It is of particular interest to note that the response decreases slightly as the operating temperature is increased (Fig. 4b). This can be attributed to the formation of metallic state in the sample above the semiconductor-metal transition temperature. Above 150°C, the increasing resistance in the metallic state may produce sufficient hindrance that prohibits any observable sensing. A maximum response value of 1.4% has been obtained for lowest detection limit of 50 ppm of $CH_4$ at lowest operating temperature of 50°C.

Response towards other interfering gases such as $NH_3$, $NO_2$ and $H_2$ has been studied. It is noteworthy that the sensor did not show any response to $NO_2$ or $H_2$. However, it showed some response to $NH_3$ which is less than 20% of the response to $CH_4$ at an operating temperature of 50°C. A possible reason for the selective response towards $CH_4$ with respect to other interfering reducing gases such as $NH_3$ and $H_2$ might be due to selective catalytic activity of $VO_2$ as mentioned in literature. One of the five valence electrons from vanadium ($4s^2 3d^3$) in $VO_2$ is free of bonding which makes it catalytically more active [21]. Oxygen molecule ionosorption ($O_2^-$) reaction is thus enhanced on $VO_2$ surface according to the well known oxygen chemisorption reaction explained by Barsan et al. [22] followed by oxidation of methane as shown in Eq. 1.

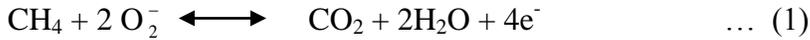

$$CH_4 + 2\,O_2^- \longleftrightarrow CO_2 + 2H_2O + 4e^- \qquad \ldots (1)$$

This effect is not observed in the case of ammonia or other gases [23]. Pristine $VO_2$ is thus shown as a promising candidate for gas sensing applications close to room temperature due to semiconducting state prevailing at near room temperatures. More in-depth mechanistic studies are required to understand the sensing behavior of $VO_2$ nano-films. The response can be enhanced by performing temperature modulated experiments across the semiconductor-metal transition region. Thermal cycling can lead to improvements in response because for each gas there is usually a point in the cycle which corresponds to a maximum in resistance-temperature profile[24]. The selectivity of $VO_2$ nanostructures to $CH_4$ close to RT can be improved by using the pattern recognition system and clustering validity measurements as mentioned in literature [25, 26].

# Conclusion

Single phase of rutile $VO_2$ films laden with nano-rod features have been successfully synthesized using two step approach involving pulsed dc-magnetron sputtering. Reversible semiconductor to metal transition is observed at ~68°C as a confirmation of the $VO_2$ phase formation. Gas sensing studies with $CH_4$ are demonstrated with $VO_2$ films showing response occurring even at low temperature of 50°C for 50 ppm of gas. Thus, $VO_2$ is a promising $CH_4$ sensing material close to room temperature.

# Acknowledgements

AKP would like to acknowledge Mrs. G. Vijayakumari for help with synthesis, Mr. S.R. Polaki for SEM experiments and Dr. A. Das for technical discussions.

# Vitae

*A. K. Prasad* received his B. Tech degree in Chemical and Electrochemical Engineering from Central Electrochemical Research Institute, Karaikudi, India in 2000 and MS degree and Ph.D. degree in Materials Science and Engineering from the State University of New York at Stony Brook, USA in 2002 and 2005 respectively. Currently he is working as Scientific Officer in Indira Gandhi Centre for Atomic Research, Kalpakkam, India. His research interests are in the area of fabrication, characterization and testing of semiconductor metal oxide based thin films and nanostructures for gas sensing applications.

*S.Amirthapandian* received his Ph.D. degree from University of Madras in 2004. Currently, he is working as a Scientific Officer at Indira Gandhi Centre for Atomic Research, in India. From 2008 to 2010, he worked as a postdoctoral researcher at University of Stuttgart in Germany. His current research interests were radiation damage in nuclear materials and micro-structural study on materials.

*S. Dhara*, MSc (1989, IIT Kharagpur), PhD (1994, NPL, New Delhi), is the Head, Nanomaterials and Sensors Section at SND, IGCAR, Kalpakkam. He is working in the field of metallic and semiconductor nanostructures for nano-sensor applications. He is an expert in the field of defect analysis in single nanostructures. He has also shown keen interest in studying plasmonic and optoelectronic properties of these nanostructures and its nanoelectronic applications.

*S. Dash* joined Indira Gandhi Centre for Atomic Research in the year 1982. He received his Ph.D in Chemistry-Materials Science from University of Madras, Chennai, India in the year 2001. Presently he is working as a Scientific Officer (H+) and heads the Thin Films and Coatings Section in Materials Science Group. He is also a Professor in Homi Bhabha National Institute. His area of expertise is coatings surface science, thin films, secondary ion mass spectrometry, tribology and equipment design for gas-solid interactions.

*N.Murali* is currently Head of the Real-Time Systems Division at the Indira Gandhi Centre for Atomic Research, Kalpakkam, India. He obtained his bachelor's degree in Electronics & Communication Engineering from PSG College of Technology, Coimbatore, India and Masters in Engineering (School of Automation) from Indian Institute of Science, Bangalore, India. His research area spans domains like Embedded & Real time systems, Distributed control system , MEMS Sensor design and characterization, Fiber Optic & Pulsating Sensors for Nuclear applications, Software engineering, GUI design using Open Source and Network protocol & Security for plant applications.

*A. K. Tyagi* obtained his Masters degree in Physics from Meerut University, Meerut, India in 1976, M.Tech degree in Solid State Materials from IIT Delhi, India in 1978 and PhD in Physics from IIT Delhi, India in 1985. He joined the Indira Gandhi Centre for Atomic Research in 1979 and is currently the head of Surface and Nanoscience Division at Materials Science Group of IGCAR, Kalpakkam, India. He is also a Professor of Physics at Homi Bhabha National Institute. His current interests include synthesis and characterization of nanomaterials and nanostructured coatings for tribology and sensor applications.

**List of Figures**

Fig 1. a) High magnification SEM image of $VO_2$ films  b) Low magnification SEM image showing regions of nano-rod bunches c) GIXRD pattern ('*' indicates substrate peaks).

Fig 2. a) TEM image of $VO_2$ nanorod bundle b) SAED pattern near the edge of nanorod bundle observed under [010] zone axis and c) HRTEM image showing grain with (201) plane.

Fig 3.   Semiconductor to Metal Transition of $VO_2$ films.

Fig 4. a) Response transient curves of $VO_2$ towards $CH_4$ at 50, 100 and 150°C (numbers beneath curves indicates concentrations), b) Comparison of sensor response (S) versus concentration at various operating temperatures.

**Fig 1**

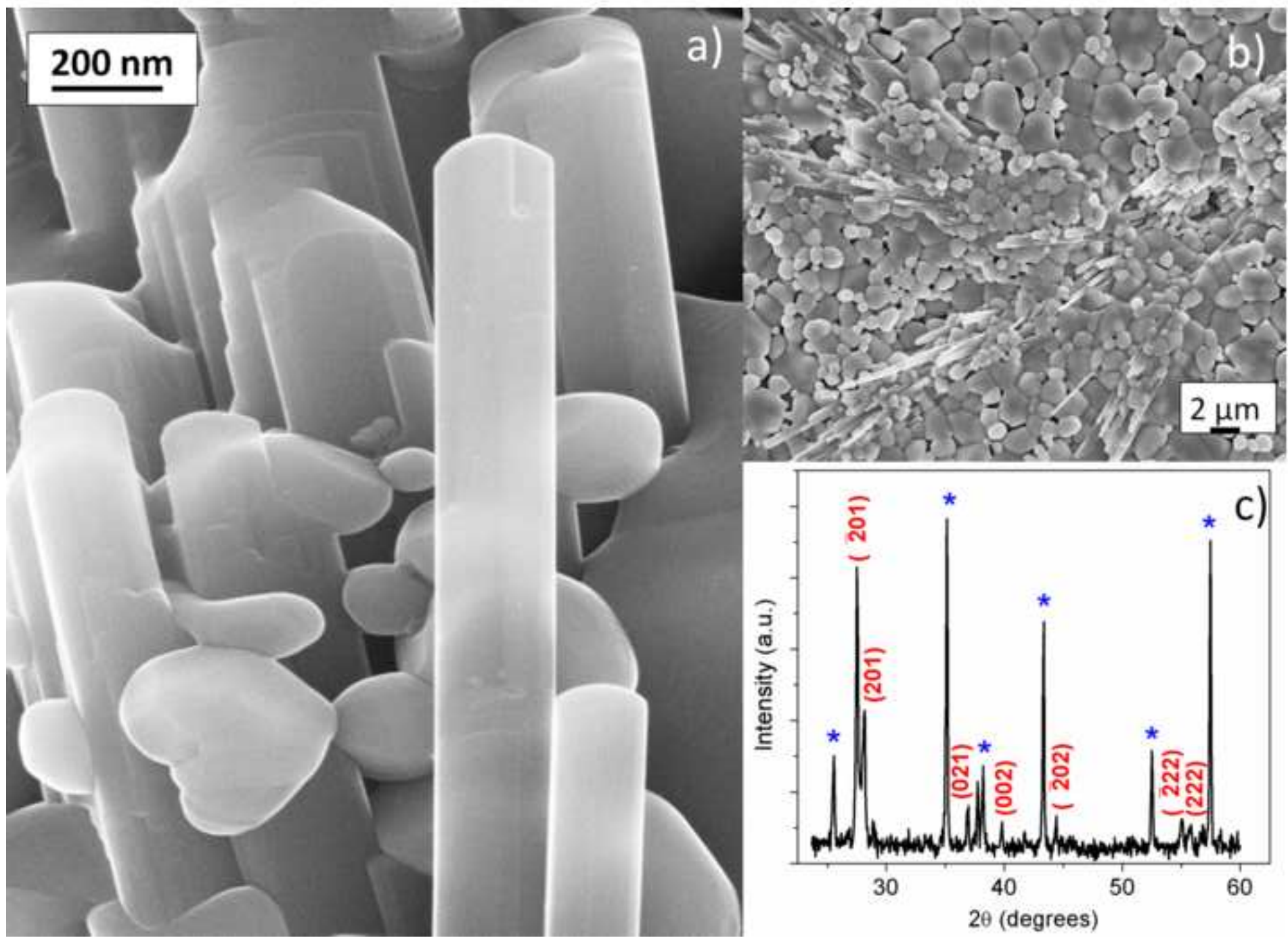

**Fig 2**

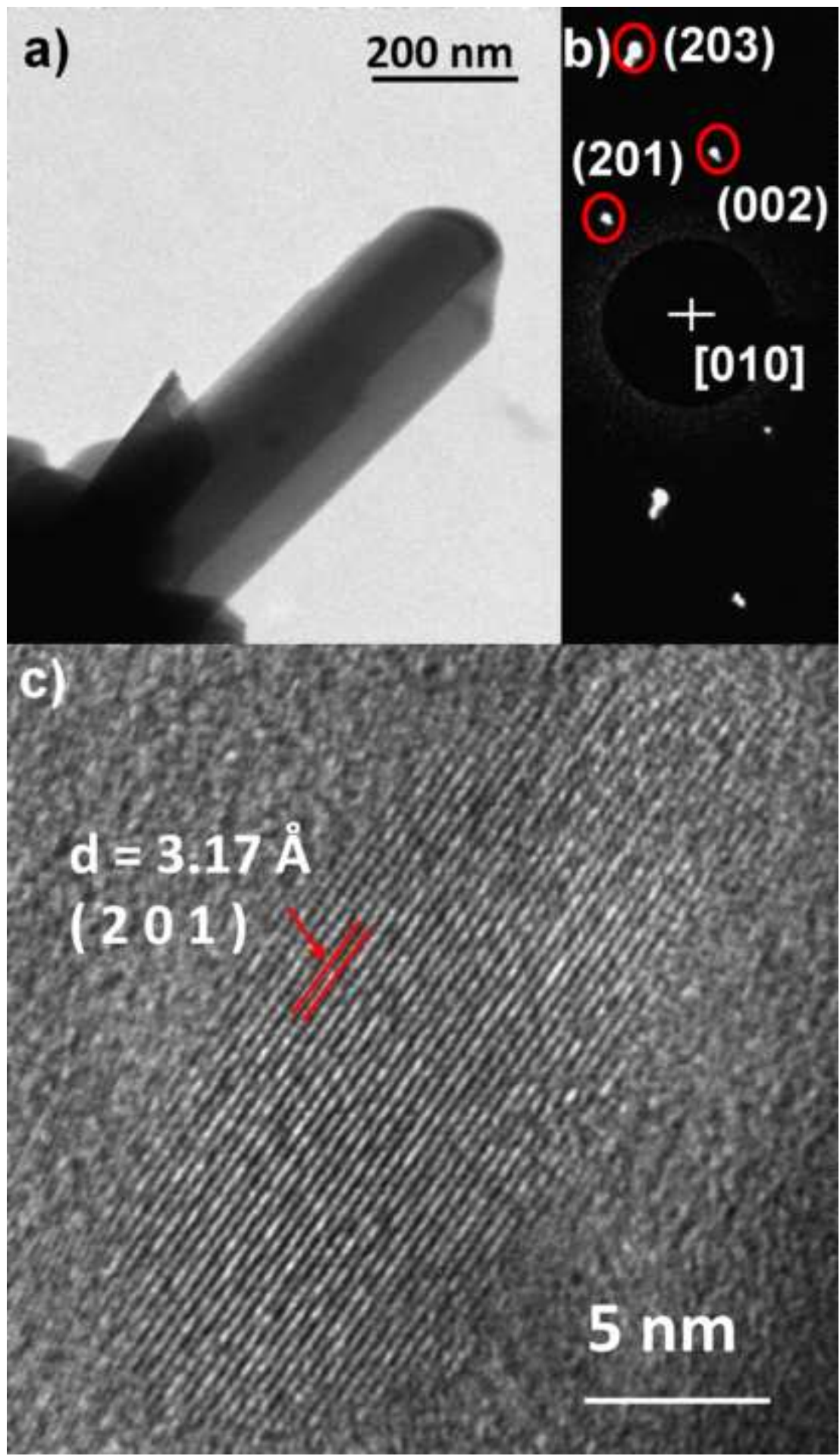

**Fig 3**

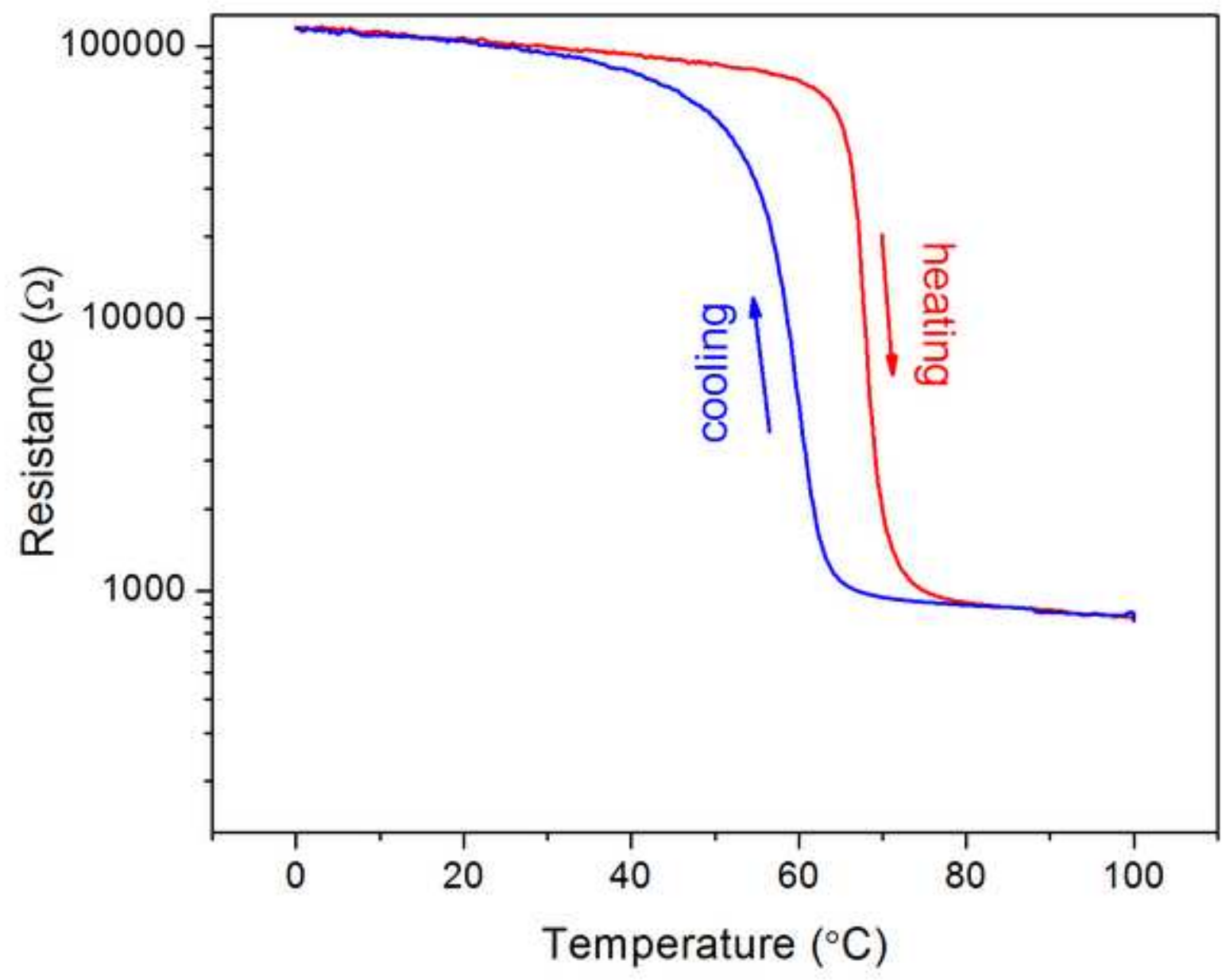

**Fig 4**

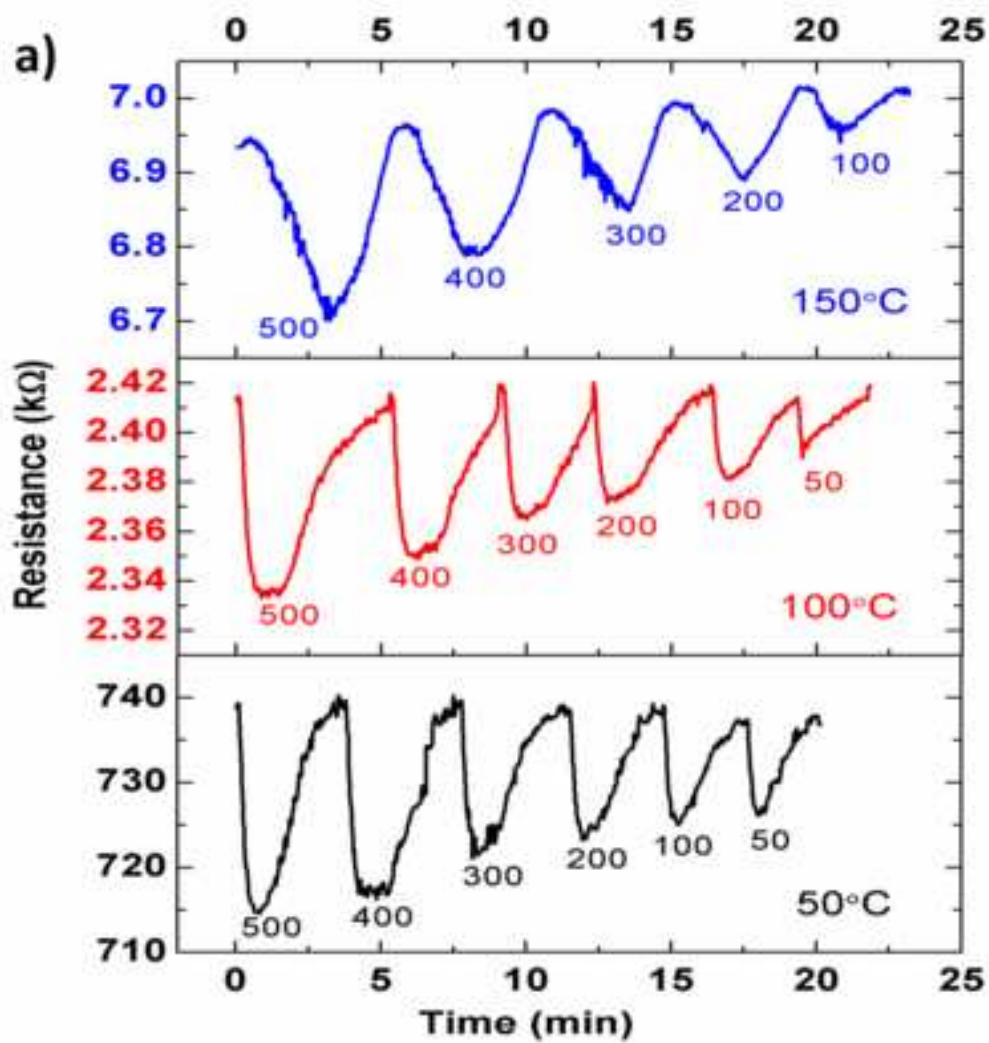

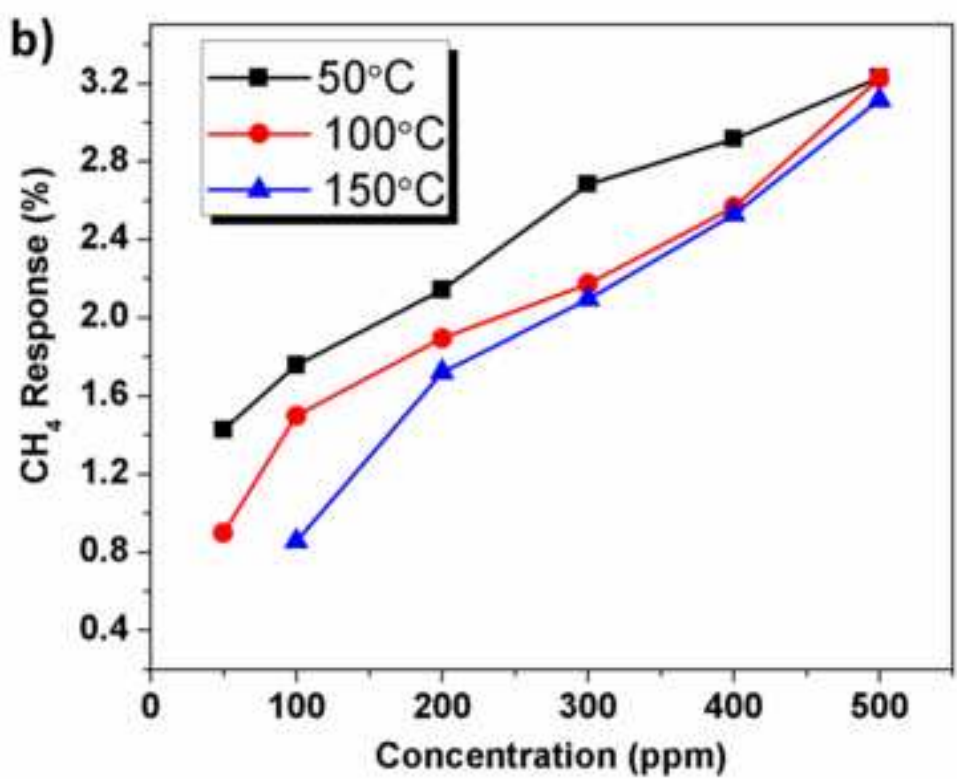